\documentclass{article}

\usepackage{amsmath}
\usepackage{graphicx}
\usepackage{url}

\newcommand{\workingnote}[1]{}        

\newcommand{\location}[1]{\ensuremath{\textit{location}({\textit{random\_ID}_{(#1)}}, t)}}
\newcommand{\randomid}[1]{\ensuremath{\textit{random\_ID}_{(#1)}}}

\title{Privacy-Protecting COVID-19 Exposure Notification Based on
  Cluster Events\footnote{This paper was presented at the \emph{NIST
      Workshop on Challenges for Digital Proximity Detection in
      Pandemics: Privacy, Accuracy, and Impact, January 28 02021}.\\ 
    Dates follow conventions of the Long Now Foundation.}  }

\author{Paul Syverson\\
  U.S. Naval Research Laboratory\\
  paul.syverson@nrl.navy.mil
}

\date{}
\begin{document}

\maketitle

\begin{abstract}
  We provide a rough sketch of a simple system design for exposure
  notification of COVID-19 infections based on copresence at cluster
  events---locations and times where a threshold number of
  tested-positive (TP) individuals were present. Unlike other designs,
  such as DP3T or the Apple-Google exposure-notification system, this
  design does not track or notify based on detecting direct proximity
  to TP individuals.

  The design makes use of existing or in-development tests for
  COVID-19 that are relatively cheap and return results in less than
  an hour, and that have high specificity but may have lower
  sensitivity.  It also uses readily available location
  tracking for mobile phones and similar devices. It reports 
  events at which TP individuals were present
  but does not link events with individuals or with
  other events in an individual's history. Participating individuals
  are notified of detected cluster events. They can
  then compare these locally to their own location
  history. Detected cluster events can be publicized
  through public channels.  Thus, individuals not participating in the
  reporting system can still be notified of exposure.

  A proper security analysis is beyond the scope of this design
  sketch.  We do, however, discuss resistance to various adversaries
  and attacks on privacy as well as false-reporting attacks.

\end{abstract}

The goal of this brief paper is to introduce the idea of, and
contextual motivation for, using mobile phone location data from
COVID-19 tested-positive (TP) individuals to identify cluster events and
then to notify people of potential exposure simply by notifying them
of cluster events. A sketch of a basic design to do this in a
privacy-protecting manner is presented. We do not discuss even
high-level particulars of necessary associated system features.

Existing privacy-protecting exposure-notification systems for COVID-19
infection generally do their detection and notification based on detected
proximity to TP
individuals~\cite{apple-google-covid,troncoso2020decentralized,private-kit}.
Location information is typically not recorded by such systems.  DP3T
intentionally ``avoids collecting location data, which is highly
sensitive and very difficult to publish in a privacy-preserving
way''~\cite{troncoso2020decentralized}. While SafePaths explores
adding GPS location information to proximity
tracing~\cite{adding-location-to-exposure-notification}, this is
considered auxiliary to the primary task of identifying individuals
who have been in close proximity to TP infectious individuals.

Deployed COVID-19 exposure-notification systems do
employ location information. For example, the
National Health Service of the United Kingdom (NHS) has set up a
system whereby venues (pubs, hairdressers, village libraries, etc.)
in England and Wales are required to obtain and post QR codes for
patrons to scan when visiting. If the venue is later determined to be a
COVID-19 hotspot, it is uploaded to a list that patrons can download
and check against visited places scanned into their
phones~\cite{nhs-qr}. This is unlike our approach in
multiple respects. It only works for fixed locations, which are
locations of establishments of specific types. It requires
participating locations to create and post QR codes. It only
works for locations that individuals have scanned into their phones.
And it identifies locations rather than events (locations at times).

Safe2~\cite{safe2} is a mobile phone app that combines self assessment
information about symptoms and proximity contact detection to provide
a ``Safe Score'' indicating risk of infection ranging from ``Healthy''
to ``Confirmed''. It also keeps track of location history, but shared
in a privacy-protecting way somewhat similar to what we suggest
below. Based on visited locations and contacts, an individual may be
classified as a likely asymptomatic carrier. Notifications will then
be sent to others who have been detected to be in close proximity to
that individual, thus decreasing their Safe Scores.

The approaches and apps mentioned above by no means constitute a
complete list, even ignoring the dynamic state of introduction and
development of new apps. Nor do we provide more than a brief note of
the features of the apps that we do mention as most relevant to or
contrasting with our approach. For example, we have not otherwise
mentioned PACT~\cite{pact} or its incorporation in SafePaths. Buchanen
et al.\ have produced a survey and analysis of privacy-preserving
COVID-19 contact tracking that was relatively comprehensive at time of
writing~\cite{review-of-pp-covid-tracing}. Landau's recent
book provides an introduction to the technology of
contact tracing and its usefulness for public health, in which she
discusses efficacy, equity, and privacy~\cite{people-count}.


For Safe2 and the other privacy-protecting systems cited above, close
proximity is determined by detected Bluetooth communication.  Dehaye
and Reardon have identified three attacks on Bluetooth-based
proximity detection in the context of COVID-19 contact tracing
apps~\cite{proximity-tracing-wpes20}. The attacks require an adversary
capable of placing a software development kit (SDK) in a moderately
successful app, but Dehaye and Reardon argue that this is much easier
than is usually assumed. See the paper for specifics. The good thing
is that we need not accurately determine how hard such attacks are or
how hard they are to counter if our exposure notification system does
not depend on phones broadcasting proximity information. And specifics
of COVID-19 infection patterns and COVID-19 testing technology may
permit simplifications of privacy-protecting notifications based on
presence at cluster events rather than directly detecting proximity to
TP individuals.

First, unlike common influenzas and other familiar diseases, COVID-19
appears to have a high degree of clustering in its dispersion and to
be spread more in events where infected individuals spend time in
close proximity to groups of others. There are other dispersion
clustering factors, such as whether gatherings are indoors and
adequacy of ventilation, and clustering also occurs around some people who
are individually linked to high number of infections.  But generally,
gatherings play a significant role in infection rates, whether or not
they are exacerbated by other
factors~\cite{kojaku2020effectiveness}. This includes both
super-spreader large gatherings for social or cultural events as well
as more moderate-sized
gatherings~\cite{sars-cov2-transmission-dynamics}.

Second, most tests for COVID-19 initially rolled out required
specialized equipment to evaluate collected samples, required days or
more to return results, were expensive, or all of the above. But
analyses indicate that faster, cheaper point-of-care tests can be more
effective at identification of infected individuals than more
sensitive tests that are
slower~\cite{Larremore2020.06.22.20136309,low-sensitivity-covid-tests}. And
U.S.\ Department of Health and Human Services in partnership with the
the U.S.\ Department of Defense is now providing rapid point-of-care
tests to communities across the United
States~\cite{hhs-rapid-point-of-care-tests}.  Similarly, under the
auspices of the World Health Organization, a global partnership has
planned to make 120 million rapid tests available in low- and
middle-income countries~\cite{who-rapid}.  Analyses of effectiveness
focus on identification and notification of infectious
individuals. But coupled with dispersion patterns, another advantage
of cheap, point-of-care tests emerges.

``[G]iven the huge numbers associated with these clusters, targeting
them would be very effective in getting our transmission numbers
down''~\cite{tufekci}. Also, identifying a cluster does not require
that everyone who was already infected at a particular event (location
and time) has tested positive or even has been tested at all. As long
as a sufficient \emph{number} of TP individuals are associated with a
given event, it is not important to identify which individuals were
present, even pseudonymously, in order for the event to constitute a
cluster.  Since cluster events are indicators of risk of infection for
all copresent at the event, informing individuals of those clustering
events at which they were present is sufficient to notify them of
potential exposure.  And, individuals can make the determination of
whether they were present at a clustering event entirely locally by
having their phones compare clustering events about which they were
notified with their location history.

Thus, we only need count the number of distinct TP individuals at an
event to identify it as a cluster event. Like other privacy protecting
COVID-19 notification systems, we can make use of anonyms (ephemeral 
pseudo-random identifiers) for each individual (where an individual is
identified with the individual's phone) at a given time. These can be
combined with location data available to the phone, e.g, by GPS and
other phone localization inputs, into an ordered pair, $( \randomid{i,t} ,
\location{i,t} \, )$. 

Phones do typically have automatic access to location data but do not
typically store location histories. Location histories are of course
frequently tracked by third parties, and apps to track location
history are readily available for the major mobile platforms. Also,
privacy-protecting tracking and storing of mobile device location data
for personal use has been studied since at least the
mid-nineties~\cite{spw97}.  As noted above, the Safe2 app specifically
stores and uses location history for COVID-19 exposure
notification. And though Safe2 relies on Bluetooth and GPS, GPS alone
can be effective at detecting useful COVID clustering
information~\cite{Serafino2020.08.12.20173476}.  The specific means of
localization is not central to our approach whether GPS, WiFi,
Bluetooth, or other background signals (such as a scanned QR-code, as
in the NHS app) as long as it is decoupled from the reporting of
location history. As with other privacy-protecting aspects not novel
in this design, we simply assume a privacy-protecting location history
system and consider particulars out of scope for this paper.

If an individual tests positive, she submits such pairs for all times
within a critical period, typically covering the maximum past interval
during which she might have become infected. (Note that backward
tracing of events appears to be more effective than forward
tracing~\cite{kojaku2020effectiveness}.  Thus it is important to trace
back to the time she might have become infected not merely the shorter
period back to the time she might have become infectious.)  Submission
can be to a decentralized repository or to a centralized repository as
long as the act of submission does not reveal association of submitted
pairs.  Given the limited goals of this paper, we simply assume such
an association-protecting submission system.  As an over-simple,
somewhat concrete example of such a system, assume a centralized
repository with submission protected by one or more
mixes~\cite{fc-handbook-survey} such that the output of the mixes is
an unordered collection of all such pairs submitted during a given
mix-system firing interval.

If notification of the results of a point-of-care test as well as
notification of exposure are automated in a COVID-19 exposure
notification app, and if adequate privacy protections are incorporated,
then individuals have self-interested incentive to use the app even
if its primary goal is to control spread. Nonetheless, as we shall
see, the system can provide useful notification of exposure even
to nonparticipants without an installed app. 

\subsubsection*{Malicious reporting of positive tests}

Another potential advantage of a cluster-event based approach is that,
even without limiting TP reporting to authorized individuals, it
provides automatic counters to the possibility that ``people may
falsely report they have been infected to cause mischief or to keep
people home in order to shut down school or even to disrupt an
election''~\cite{cranor-covid}. An individual falsely reporting a
positive test cannot easily create such a result because they are
unlikely to be the one to transition a location and time across the
threshold of counting as a cluster event. And they cannot report at
all for a location and time unless they were present at that
event. More significant coordinated copresence or system hacking of
location reporting would be required. And if coordinated copresence is
the mechanism, then it may be that innocents who are notified because
they were present at that cluster event \emph{should} seek further
testing and curtailing of social interactions.

Our cluster-event design thus automatically prevents, e.g., a student
who is worried about his exam next week and anonymously reports a
positive test from thereby causing his school to shut down or his
whole chemistry class to be forced into quarantine. A fan of
one sports team cannot force a rival team into quarantine, etc.

A more substantial adversary, such as a nation-state, might be able to
hack location histories for a phone, create sybils of phones at a
location, etc.  This could support an attack at a larger scale to
disrupt a critical operation or degrade the availability of essential
infrastructure or emergency personnel. Since the proposed design is
meant to leverage rapid, point-of-care tests, it is also conducive to
requiring input from authorized testing personnel at point of care to
permit reporting a positive test. For example, an authorized TP code
tied to a unique anonym bound to the identity of the mobile phone
present when an individual is given the test could be sent at the same
time test results are comunicated to that phone, possibly even after
the individual has left the point of care. And reporting of recent
location history for that phone would require this
authorization. Specifics of what capabilities such more significant
adversaries might have and how all this would work is beyond the scope
of this paper. We simply note here that the cluster-event approach
remains compatible with requiring authorized parties to confirm TP
status in order for reporting to occur.

Obviously there is a tension between authorization mitigations against
more powerful false reporting attacks and the ease and effectiveness
of participation in the reporting and notification system. For
example, without an authorization requirement, simple tests
requiring no expertise to administer or evaluate can be made
available and incorporated in the exposure notification system without
an authorized-testing-personnel bottleneck.  And a powerful adversary might
include many other elements beyond our scope, such as compromising
trusted authorization individuals, or the systems used for authorization,
or the physical tests themselves. Lesser adversaries described above
are still countered by the proposed system even without the authorization
component. This is another advantage over any system that inherently
has a trusted-authorization bottleneck.

\subsubsection*{Cluster event criteria}

Submitted anonym pairs can be clustered into events according to
latest understanding of what constitutes sufficient proximity in
space and time to indicate risk of exposure likely enough to merit
notification. Such clustering can be based simply on the number of
copresent TP individuals. Further privacy preserving measures exist
that would only reveal an uploaded location and time if a cluster
event occurs there. Indeed there are deployed systems for large-scale
gathering of data and release of associated statistics with guarantees
of differential privacy~\cite{prochlo}. 

With small thresholds of cluster events, however, even if the system
only reveals locations if there is a cluster event, there may be a
tension between revealing cluster events in a differentially private
way while also not significantly affecting the rate of false negatives
about clusters. This may not be a significant limitation, however.
Without differential privacy, in principle an adversary could submit a
false positive result to enhance tracking of a TP individual. But this
is not trivial. If, for example, the adversary had a device that could
report from the individual's location, easier tracking is
available. So we would need to assuming knowledge of suspected
trajectories followed by the targeted TP individual, and a hack of the
location reporting system that doesn't also make more straightforward
tracking possible.  In that case, the adversary could report TP
locations and times along the suspected route to look for instances
where the adversary's report caused a cluster event.  If so, and
ignoring the entrance of other TP individuals causing the transition,
the target's presence could have caused locations-times to be at one
less than the cluster-reporting threshold.  These issues are also
beyond the scope of this paper.

Exogenous information, if available, could affect clustering
classification. For example, a location may have been separately
identified as that of a salon, restaurant, house of worship, etc. This
can affect what constitutes a notification-worthy cluster: two TP
individuals copresent on a street corner in a city may be too low a
threshold to classify this as a cluster event meriting notification,
two TP individuals copresent in a poorly ventilated small salon might
be a reasonable threshold, however.

Conversely, cluster locations that are identified simply by the number
of copresent TP individuals but otherwise unknown and that are also
responsible for large or repeated cluster events might be flagged for
public health officials to investigate what is at that location and
what sort of people are going there for what purposes.

On the other hand, unlike direct contact detecting approaches, since
notification is only of cluster events, this approach cannot detect
and therefore cannot notify people if they are exposed to a single
copresent individual. As noted above, spread of COVID-19 is primarily
through such clustering events, however.  And this approach will also
provide notification in a circumstance where an individual might never
have come in close contact to anyone who tested positive (who is also
participating in the detection/notification system) but was present at
an event where multiple TP individuals were detected.  So it not clear
which approach is more likely to result in notification. Further,
individuals from communities which have historically experienced
disproportionate
negative impacts of public health crises might be disinclined to
participate in a contact tracing
system~\cite{contact-tracing-equity}. Even if a non-participant, they
might nonetheless self-interestedly check a public website for cluster
events and act if they notice a significant cluster at an event where
they (or loved ones) were known to be present. Recall that in any
case, the primary goal is to produce notifications that lead to
significant reduction in spread rather than to guarantee an individual
of notification upon any exposure.

Relatedly, the approach allows different criteria for distance in
making a clustering determination. As an oversimplified illustration,
suppose a 2000 $m^2$ area is comprised of dozens of TP individuals
where all individuals on the perimeter are within 2 meters of another
TP individual for several minutes. The relevant cluster event for
notification should probably cover the entire area at that time even
if portions of the center are 10 meters from the nearest TP
individual---unless there is specific additional information that
would exclude particular subareas. It may be likewise reasonable to
include a larger area than just 2 meters beyond the perimeter for such
a large cluster.

Another potential limitation of this approach is that it identifies a
cluster as at one (possibly extended) location in a relatively brief
time interval. It does not directly have a means to distinguish
extended presence at a location of multiple TP individuals with few
comings and goings over a period of hours (for example at a social
gathering) versus a roughly persistent total of TP individuals
resulting from regular turnover (for example, at a transit hub or
commercial service establishment). This could matter for time of
exposure to a particular individual as an indicator of infection risk
likelihood.  On the other hand, whether a notified individual learns
of exposure at multiple cluster events at the same location based on
either of these scenarios may not matter for the recommended course of
action.  Notified individuals are also more likely to know the
circumstances of a particular succession of notified cluster
events. And if they know the circumstances, they may be more prepared
to notify others personally known to be associated with those
circumstances, whether or not those others participate in or monitor
the notification system.

Relatedly, there is no distinction made by the basic system of
multiple exposures at a succession of locations versus prolonged
exposure, e.g., during a shared ride in a car or mass transit vehicle.
Again, it might not make a practical difference for purposes of
the system whether this is identified as a single prolonged cluster event
or succession of many shorter ones.

\subsection*{High-level Design Summary}

\begin{enumerate}
  \setlength{\itemsep}{.5mm}
\item Individual tests positive for COVID-19 and enters this in the
  notification app on their phone (alternatively notification of a
  positive result might be automated in a phone app and/or require
  input from a trusted testing official).
\item Individual's phone app prepares a historical list of locations
  and times since the beginning of the critical period based on time
  of testing, each paired with a different random identifier.
\item Phone submits the list of pairs through a privacy preserving system.
\item System receiving these pairs from all reporting phones clusters
  them into events at which multiple TP individuals are present.
\item System pushes notifications of all cluster events to participants'
phones and/or posts these to a publicly accessible location, e.g., a website.
\item Notified individuals learn of exposures at cluster events. 
\end{enumerate}

The purpose of this brief paper is to introduce a novel concept of
COVID-19 exposure notification based on creation of clustering events
of which individuals are notified.  We have simply assumed adequately
privacy preserving systems for data gathering, processing, and
publishing.  And there are existing systems that make this assumption
plausible, some of which we have cited. Details matter, however, to
the security, scalability, practicality, and usability of the overall
system. Some of the details we have ignored include how submission of
pairs unlinks submitting individuals from locations and histories of
locations, the number of submitting individuals (in a region, during
an interval), resistance to de-anonymization of location histories
from plausible travel paths given a collection of location-time pairs,
how long various data and information are held, etc. Clustering
algorithms and clustering criteria are central to the cost, basic
viability, and properties provided by this approach, but we have
simply assumed these will be selected to function as needed. Cognizant
of these assumptions, we have nonetheless identified a number of
high-level properties of this approach, which we now summarize.

\subsection*{Design Features}

\begin{itemize}

\item Does not depend on TP individual's phone to have Bluetooth
  turned on in order to provide inputs.

\item Does not depend on potentially exposed individuals to have
  Bluetooth on in order for system to find indications of exposure.

\item Does not depend on direct physical interaction (Bluetooth or
  otherwise) between phones of TP individuals and exposed individuals.

\item Exposure contact distance parameters do not depend on limitations of
  Bluetooth communication parameters.

\item Does depend on availability of phone location history of
  reporting TP individuals.

\item Does depend on phone location history (or human memory or\ldots)
  for individuals to effectively make use of cluster event
  notifications or postings.

\item Does reveal locations and times of cluster events, even small
  ones, which may identify particular households where there are
  multiple infections or that some sort of meeting took place at an
  otherwise nondescript location.

\item Leverages fast test results, even if tests have only modest sensitivity.

\item Depends on approximate numbers of TP individuals in cluster
  events rather than specific contacts with TP individuals.

\item Cannot detect exposure to TP individuals outside of cluster events.

\item Can notify of exposure even without close contact to a TP individual.

\item Can notify of exposure even without participation as reporting
  individual.

\item Counters false TP reporting \emph{and} tolerates false TP reporting.

\item Compatible with using trusted authorizations to counter false TP
  reporting by stronger adversaries.

\end{itemize}

\section*{Acknowledgements}
Thanks to Aaron Johnson, Rob Jansen, Matt Traudt, and Ryan Wails for
helpful discussions, and to Paul-Olivier Dehaye and Susan \mbox{Landau} for
multiple helpful comments and discussions.

\end{document}